\documentclass[pre,aps,twocolumn,]{revtex4}
\usepackage{graphicx}
\usepackage{dcolumn}
\usepackage{bm}
\usepackage{amssymb}
\usepackage{pifont}
\usepackage{amsmath}
\usepackage{times}

\usepackage{color}

\begin{document}

\title{Modelling Trading Networks and the Role of Trust}

\author{Rafael A. Barrio and \'Elfego Ruiz-Guti\'errez}

\affiliation{Instituto de F\'isica, Universidad Nacional Aut\'onoma de M\'exico, 20-364, 01000 M\'exico DF, M\'exico}

\author{Tzipe Govezensky}

\affiliation{Instituto de Investigaciones Biom\'edicas, Universidad Nacional Aut\'onoma de M\'exico, 04510 M\'exico DF, M\'exico}

\author{Kimmo K. Kaski}

\affiliation{Department of Computer Science, Aalto University School of Science, FI-00076 AALTO, Finland}

    \begin{abstract}
		We present a simple dynamical model for describing trading interactions between agents in a social network by considering only two dynamical variables, namely money and goods or services, that are assumed conserved over the whole time span of the agents' trading transactions. A key feature of the model is that agent-to-agent transactions are governed by the price in units of money per goods, which is dynamically changing, and by a trust variable, which is related to the trading history of each agent. All agents are able to sell or buy, and the decision to do either has to do with the level of trust the buyer has in the seller, the price of the goods and the amount of money and goods at the disposal of the buyer. Here we show the results of extensive numerical calculations under various initial conditions in a random network of agents and compare the results with the available related data. In most cases the agreement between the model results and real data turns out to be fairly good, which allow us to draw some general conclusions as how different trading strategies could affect the distribution of wealth in different kinds of societies. \\

    \end{abstract}

\maketitle

\section{Introduction}

In human societies social life consists of the flow and exchange of norms, values, ideas, goods as well as other social and cultural resources, which are channeled through a network of interconnections. In all the social relations between people {\em trust} is a fundamental component~\cite{gamb}, such that the quality of the dyadic relationships reflects the level of trust between them. From the personal perspective social networks can be considered structured in a series of layers whose sizes are determined by person's cognitive constraints and frequency and quality of interactions~\cite{dunbar}, which in turn correlate closely with the level of trust that the dyad of individuals share. As one moves from the inner to the outer layers of an individual's social network, emotional closeness diminishes, as does trust. Despite its key role in economics, sociology, and social psychology, the detailed psychological and social mechanisms that underpin trust remain open. In order to provide a systematic framework to understand the role of trust, one needs to create metrics or quantifiable measures as well as models for describing plausible mechanisms producing complex emergent effects due to social interactions of the people in an interconnected societal structure.

One example of such social interaction phenomena, in which trust plays an important role, is trading between buyers and sellers. Such an economic process is influenced by many apparently disconnected factors, which make it challenging to devise a model that takes them into account. Therefore, models that have been proposed, necessarily select a subset of factors considered important for the phenomena to be described. For instance, there are studies of income and wealth distribution~\cite{sinha,wd}, using gas like models~\cite{bikas2007}, life-cycle models~\cite{wang}, game models~\cite{bikas2015}, and so on. For a review of various agent based models we refer to~\cite{sornette}. In addition, we note that detailed studies of empirical data and analysis of the distribution functions~\cite{yuqing2007} seem to lend strong support in favour of gas-like models for describing economic trading exchanges.

In order to consider the role of trust in trading relations we focus on the simplest possible situation in which trust clearly plays a definite role. This is the case of trading goods or services for money through dyadic interactions or exchange, which takes place either as a directional flow of resources from one individual to another individual or vice versa. When an agent is buying, trust plays a role, as people prefer to buy from a reliable and reputable selling agent, i.e. agent they trust. It should be noted that the dyadic relationship does not have to be symmetric, i.e. a seller does not need to trust the buyer. A key ingredient in the trading interactions is profit that an agent makes when providing goods or services, and it can realistically be assumed that a seller wants to put the highest possible price to its goods, while the buyer tends to perform operations with agents offering a low price.

In this study we propose an agent based "gas-like" model to take into account the above mentioned important features of trading. The model describes dyadic transactions between agents in a random network. The amount of goods and money are considered conserved in time, but the price of goods and trust, we measure as reputation, vary according to the specific situation in which trade is made. In section~\ref{model} we describe the model and set up the dynamic equations of the system. In section~\ref{results}  we present the results of extensive numerical calculations and explore their dependence on the parameters of the model. Here we also compare our numerical results with available real data and discuss the predictions of the model as well as possible extensions to it. Finally, in section~\ref{conclusion} we conclude by making remarks concerning the role of trust in trade and social relations.

\section{The model}
\label{model}

\subsection{The basic model}
First we introduce the basic model, which describes the dynamic development of a random network of $N$ agents such that the state of agent $i$ is defined by two time-dependent state variables, $(x_i,y_i)$, where $x_i$ stands for the amount of money and $y_i$ for the amount of goods or services. The pairwise connectivities between agents in the network are described by $N\times N$ adjacency matrix $C$.
It is necessary to distinguish the direction of the flow of goods and money in the network, since agent $i$ could buy from agent $j$, or vice versa. At time $t=0$ we define two symmetric matrices, $A(t=0)$ and $B(t=0)$, with an average of $Z/2$ random entries per row, for the flow of money or goods, respectively. Then the adjacency matrix is simply $C=A+B$, and $Z$ stands for the mean degree.

 The elements of $A(t)$  and $B(t)$ are defined as the normalised probabilities of transactions per unit time $\alpha$ and $\beta$, respectively and they could become asymmetric. These matrices represent the buying or selling transactions, according to the individual agent instantaneous situation.

The dynamic equations for the state variables $x$ (money) and $y$ (goods) initialised randomly $\in [ 0,1]$ are,
\begin{subequations}\label{nm}
\begin{align}  \frac{\mathrm{d}x_{i}}{\mathrm{d}t} & = \sum_{j} \left[- x_{i}\beta_{ij}-s_jy_j \alpha_{ji}+ x_{j} \beta_{ji}  + s_i y_i\alpha_{ij},\right] \label{second} \\
 \frac{\mathrm{d}y_{i}}{\mathrm{d}t} & = \sum_{j} \left[ \frac{x_i \beta_{ij}}{s_j} +y_j\alpha_{ji}- \frac{x_{j}\beta_{ji}}{s_i}-y_i\alpha_{ij}\right].\label{third}
 \end{align}
\end{subequations}
where $s_i$ is the price of the goods as decided by seller $i$, and its value depends on time. In both Eqs.~(\ref{nm}) the first and second terms on the right represent the transactions in which agent $i$ is buying goods from agent $j$. Note that there is an outflow of money (negative $\beta_{ij}$) and an inflow of goods (negative $\alpha_{ji}$). The third and last terms represent selling goods to $j$.  Observe that we could simply use the first and third terms in Eq.~(\ref{second}), to represent the exchange of money, or use the second and fourth terms, if the exchanged money is represented as its transaction value $sy$ ($s$ has units of money per unit good). The same reasoning is applied to Eq.~(\ref{third}). We preferred to keep the equations in a symmetric form as regards to goods and money.

The elements of the matrices $\alpha$ and $\beta$ in the equations represent the  proportion of money and goods, respectively, that an agent is distributing amongst its links in a given transaction. Therefore, the simplest expressions for the corresponding transaction coefficients are linear functions of the respective variables, thus we propose,

 \begin{equation}
 \begin{aligned}
 \alpha_{ij}(t) &= \frac{1}{\Delta t}\frac{x_{j} \Theta(s_{i}y_i - x_{j} )}{\sum_j x_j\Theta(s_{i}y_i - x_{j} )},\\
\beta_{ij}(t)&=\frac{1}{\Delta t}\frac{y_{j} \Theta(x_i-s_{j}y_j)}{\sum_j y_j \Theta(x_i-s_{j}y_j)},
  \end{aligned}
  \label{ab}
\end{equation}
provided $\sum_i A_{ij}\ne0$ for $\alpha$ and $\sum_i B_{ij}\ne0$ for $\beta$, respectively. The unit time $\Delta t$ could conveniently be taken as one. The reason to include the Heaviside functions $\Theta$ is that agent $i$ buys goods from agent $j$ only if it has enough money to pay for the price agent $j$ is asking for its goods (in $\beta_{ij}$) or if agent $j$ has enough goods to satisfy agent $i$ (in $\alpha_{ji}$). By the same token, agent $i$ cannot sell to agent $j$ if it has not enough goods (as in $\alpha_{ij}$) or if the buyer has not enough money (as in $\beta_{ji}$).

The diagonal elements of these matrices represent the quantity of money or goods that are not used in the transaction. If they are set to zero, all the money that an agent has is used in all the transactions, meaning that there are no savings.  The elements of these matrices cannot be negative, hence, whenever an element is about to become negative, it should be set to zero, which is equivalent to say that the link between agents  is irrelevant for that particular transaction.This amounts to reshaping the trading network.

These dynamical equations constitute the basic model for trading transactions. It should be noted  that the model conserves the amount of goods $(\sum_i x_i)$ and money $(\sum_i y_i)$. This restriction could be easily relaxed, if needed, by adding up sources and sinks to the equations. We decided to keep the conserved version for simplicity.

\subsection{Trust is built from reputation}

As emphasised above trust plays a key role in all social and trading transactions. In order to include trust into the transactions, we need first to assume that trust is something that you either gain or lose with time.
Therefore, it has to be related to a quantity that measures the performance of agents as traders, which we here assume to be {\it reputation} $R_i(t)$. Accordingly, we write
\begin{equation}
  \begin{aligned}
    R_i(t) & = \frac{ \mathcal{R}_i(t) }{\max_i \{ \mathcal{R} (t) \}}, \qquad \text{ where } \\
    \mathcal{R}_i (t) & = \int_{0}^t \sum_{j=1}^N \left[ \alpha_{ij}(t') + \alpha_{ji}(t') + \beta_{ij}(t') + \beta_{ji}(t') \right] \mathrm{d} t'.
  \end{aligned}
\label{R}
\end{equation}

Observe that the numerator of Eq.~(\ref{R}) at a given time increases with the number of successful transactions, since the elements of the trading matrices $\alpha$ and $\beta$ are positive definite. In order to maintain the scale of variation of all the variables between zero and one, we normalise the variable $R$ by dividing it with the maximum value of $R$, encountered in the network. Note that the reputation of a given agent could decrease in time because of the time dependent  normalisation.

Let us now introduce a non-symmetrical matrix $F$, whose elements $F_{ij}$ regulate the amount of goods and money in each transaction between a pair of agents. The entries of the matrix $F$ range from zero (no transaction possible) to unity (larger amount of goods and money exchanged). We assume that agent $i$ prefers to buy if the value of the goods in its possession is smaller than the amount of money it has, and favours selling otherwise, in order to maintain a balanced stock of money and goods, and the same for agent $j$. To reflect this, one could define the matrix elements of $F$ as follows,

\begin{equation}
 F_{ij}(t) = R_j \Theta(\hat{s}(t) y_i(t)-x_i(t)).
  \label{F1}
\end{equation}
where $\hat{s}(t)$ is the average price over the whole network of the goods at time $t$, which could be very different from the price the individual $i$ is pricing its goods. Here it is assumed that agent $i$ is buying from agent $j$, and therefore the reputation of agent $j$ is the factor that matters. Trust on sellers when buying is important because one buys from a reliable agents only. On the contrary,  when selling one is happy to do it to whomever is able to pay, independently of the reputation of the buyer.
Note that the transpose $F_{ji}$, should be used on the transactions in which agent $i$ sells goods, in which case the reputation of agent $i$ matters.

Taking this new feature into account, the dynamical equations of our model read as follows,

\begin{equation}
\begin{aligned}
  \frac{\mathrm{d}x_{i}}{\mathrm{d}t} = \sum_{j} & \left[ - x_{i}\beta_{ij}F_{ij} - s_jy_j \alpha_{ji}F_{ij} \right. \\
  & \qquad \left. + x_{j} \beta_{ji}F_{ji}  + s_i y_i\alpha_{ij}F_{ji},\right], \\
  & \\
  \frac{\mathrm{d}y_{i}}{\mathrm{d}t} = \sum_{j} & \left[ \frac{x_i \beta_{ij}F_{ij}}{s_j} +y_j\alpha_{ji}F_{ij} \right. \\
  & \qquad \left. - \frac{x_{j}\beta_{ji}F_{ji}}{s_i}-y_i\alpha_{ij}F_{ji}\right].
 \end{aligned}
 \label{nmF}
\end{equation}

Notice that these model equations reduce to those of the basic model (Eqs.~(\ref{nm})) if one eliminates the effect of trust by setting $F_{ij}=F_{ji},=1, \; \forall i,j$. In this sense, the effect of trust is to regulate the amount of goods and money exchanged in each transaction, but also in deciding if one sells or buys.

\subsection{Evolution of prices $s(t)$}
The main quantity that regulates the dynamics of goods and money in our trading model  is the  time dependence of the price of the goods. Therefore, one needs  to suggest a mechanism by which an agent modifies the price of its goods. We assume that in general people would like to sell at the highest possible price, but we need to consider to what extend it is limited by the agent's reputation.
If the agent is reputed to be a successful trader, it could increase its prices by a large amount, provided that the price the agent $i$ is using is smaller than the average price the agents connected to it are offering, otherwise lowering the value of $s_i$ is a way of attracting more transactions. In a mathematical language, one can represent this situation by the following coupled first order equations,

\begin{equation}
\label{sd1}
\begin{aligned}
\frac{\mathrm{d} \tau_i}{\mathrm{d}t} & = - \frac{1}{t_0} R_i(t) \left[ s_i(t) - \frac{1}{k_i} \sum_{j=1}^{k_i}s_j(t) \right], \\
 & = -\frac{1}{t_0} R_i(t) \left[ s_i(t) - s^k_i(t) \right], \\
\frac{\mathrm{d} s_i}{\mathrm{d}t} & = \frac{1}{t_0} \left[ R_i \tau_i(t) - g s_i(t) \right],
\end{aligned}
\end{equation}
where $\tau_i$ is a variable that enables agent $i$ to decide how to modify its prices according to trust and the information available to it. The parameter $t_0$  is the time scale for price changes and $g$ is a constant that represents reluctance for the agent to lower the prices, which for simplicity is assumed to be very small and the same for all the agents.

We can eliminate the variable $\tau$ from this equations and obtain a second order equation for $s$, which we identify as a set of damped harmonic oscillators, which admits an
analytical solution, namely,

\begin{equation}
\label{dho}
s_i(t)=s_i(0)e^{-\gamma t}\left[\cos (\omega_i t)-\frac{\gamma}{\omega_i}\sin (\omega_i t) \right] + s_i^k(t),
\end{equation}
where $\gamma=g/2$ and $\omega_i=\sqrt{(4R_i^2-g^2)}/2$. Notice that this solution is approximate and valid only when the prices do not change appreciably within a time step $\mathrm{d}t$, which means that $t_0$ is large and $g$ is small. In any case one could find the solution of this non-linear system of coupled oscillators numerically. The solutions of Eqs.~(\ref{sd1}) produce an apparently chaotic behaviour for $s(t)$ showing qualitatively similar to the price variations of real markets (there are even some models of the stock market that use damped oscillators to analyse the price variation~\cite{sor}).

\section{Results.}
\label{results}
The numerical calculations were made in a random network of 500 agents with the average degree of 12, using a simple Euler method with random initial conditions for $x$, and $y \, \in [0, 1]$. The appropriate step size for convergence was found to be $\mathrm{d}t = 0.05$, reached within 200000 iterations. At time $t = 0$ we set $s(0)$ randomly using a flat distribution with the mean $\bar{s}=1.2$ and the width $\mathrm{d}s = 0.4$, and set a time scale $t_0 = 400$ and $g = \mathrm{d}t / t_0$. For this $t_0$ the system reaches dynamical equilibrium within the running time. The diagonal elements of the trading matrices $A(t)$ and $B(t)$ were set to zero.

\begin{figure}[ht!]
\begin{center}
\includegraphics[width=6cm]{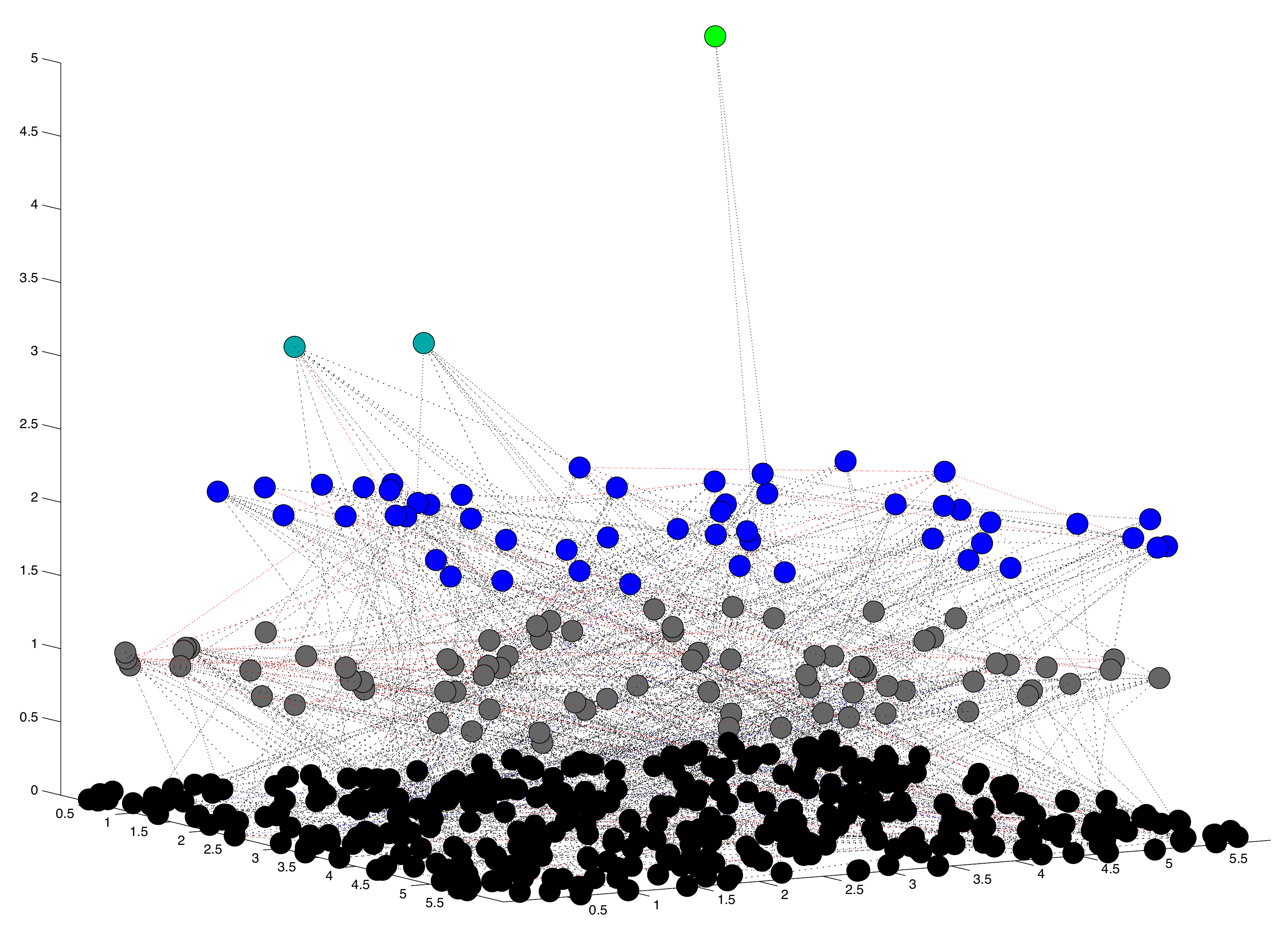}
\caption{Final configuration of an active trading network of buyers and sellers when trust is included in trading transactions. The colour code for the links between agents is as follows: Black $\alpha\ne 0$ and $\beta \ne0$, Red $\alpha\ne 0$ and $\beta =0$, and Blue $\alpha=0$ and $\beta \ne0$. The colour of the nodes is chosen according to the wealth ($w_i=x_i+y_i)$ such that light green, red, dark green, blue, grey, black are in decreasing order of $w$.}
 \label{fig1}
\end{center}
\end{figure}

In Fig.~\ref{fig1} we depict the structure of the final network by showing only the active links, where the height of the 3D plot is stratified according to the wealth of the agent $i$, i.e. $w_i = x_i + y_i$. Here it is seen that there are poor agents and only a few very rich ones. In addition we observed in numerous calculations that if the width of the distribution of prices is increased, the rich become richer and the number of poor agents increases.

\begin{figure}[ht!]
\begin{center}
\includegraphics[width=\columnwidth]{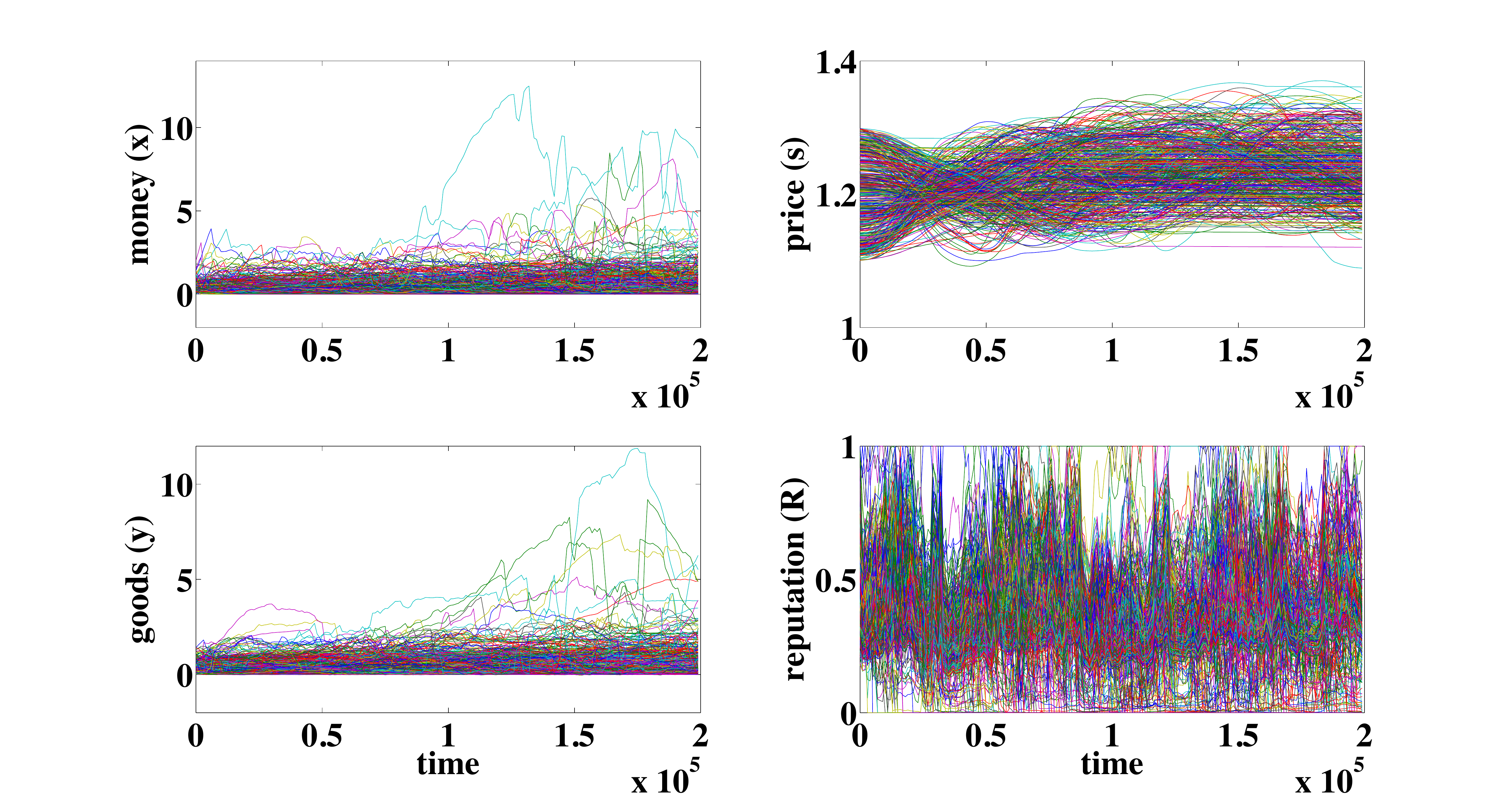}
\caption{The time histories of the variables for amounts of traders' or agents' money and goods, and prices of goods and sellers' reputations.}
 \label{fig2}
\end{center}
\end{figure}

In Fig.~\ref{fig2} we show the time histories of the variables for the typical calculation of the previous figure. Note that most agents have converged to certain amounts of money and goods, but the ones who posses more wealth are prevented from taking it all, which is an effect of trust. Also observe oscillations of prices and apparently chaotic behaviour of trust. With the chosen value of $g$ (reluctance for the agent to lower the prices) the average price is maintained around a constant value, but with increasing $g$ the average price diminishes, because one is less reluctant to lower its prices. If all prices are set to the same value, the dynamical behaviour of $R$ turns out to be less chaotic, and shows less variations.

\begin{figure}[ht!]
\begin{center}
\includegraphics[width=\columnwidth]{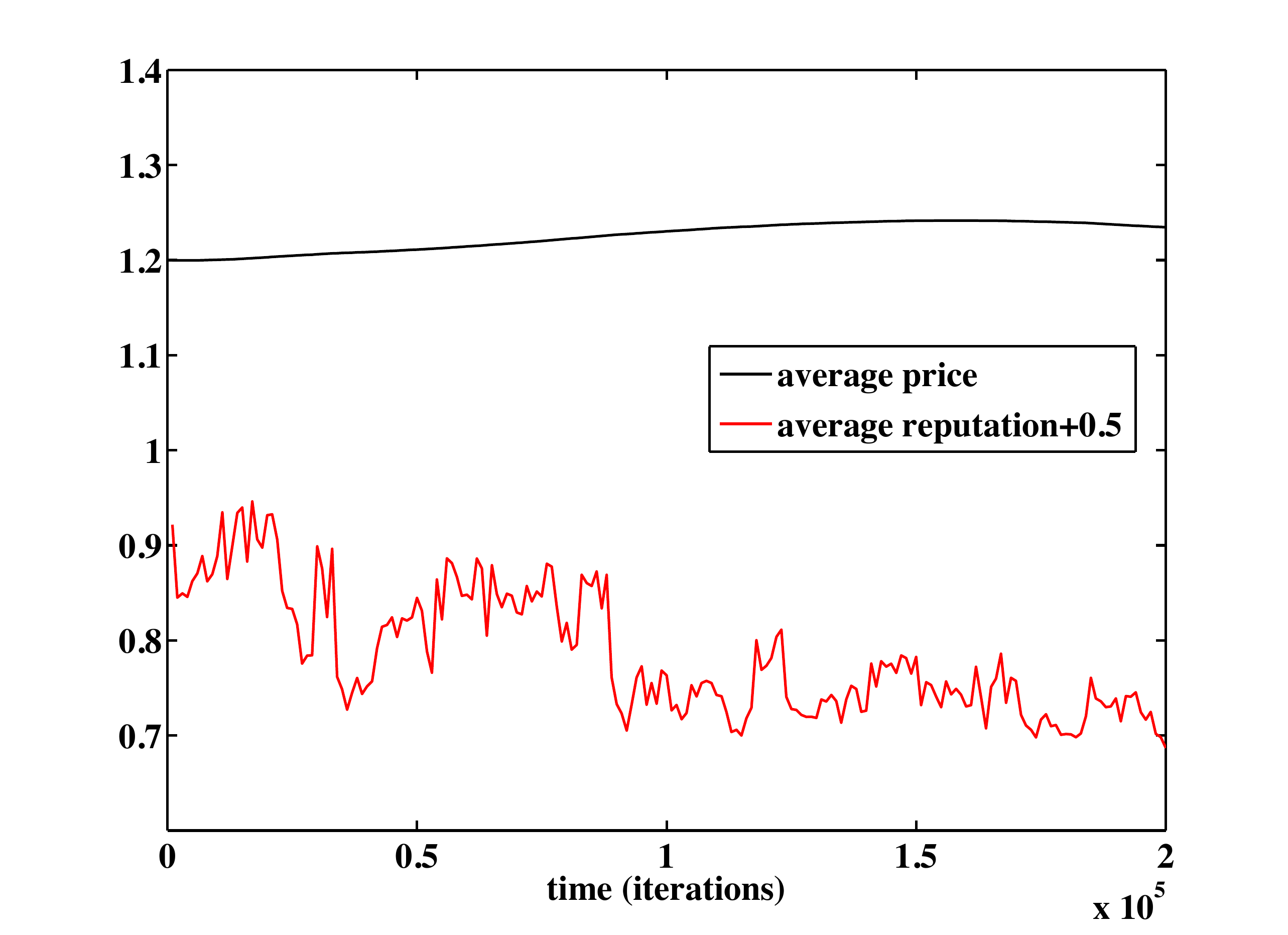}
\caption{The time histories of the average price of goods and reputation of sellers in the network.}
 \label{fig3}
\end{center}
\end{figure}

In Fig.~\ref{fig3} the behaviour of the average reputation of sellers and price of goods are depicted. Here we see that the mean value of reputation of agents decreases with time, which reflects the fact that there are less successful transactions, while the reputation of some selected individuals increases with time.

\begin{figure}[ht!]
\begin{center}
\includegraphics[width=\columnwidth]{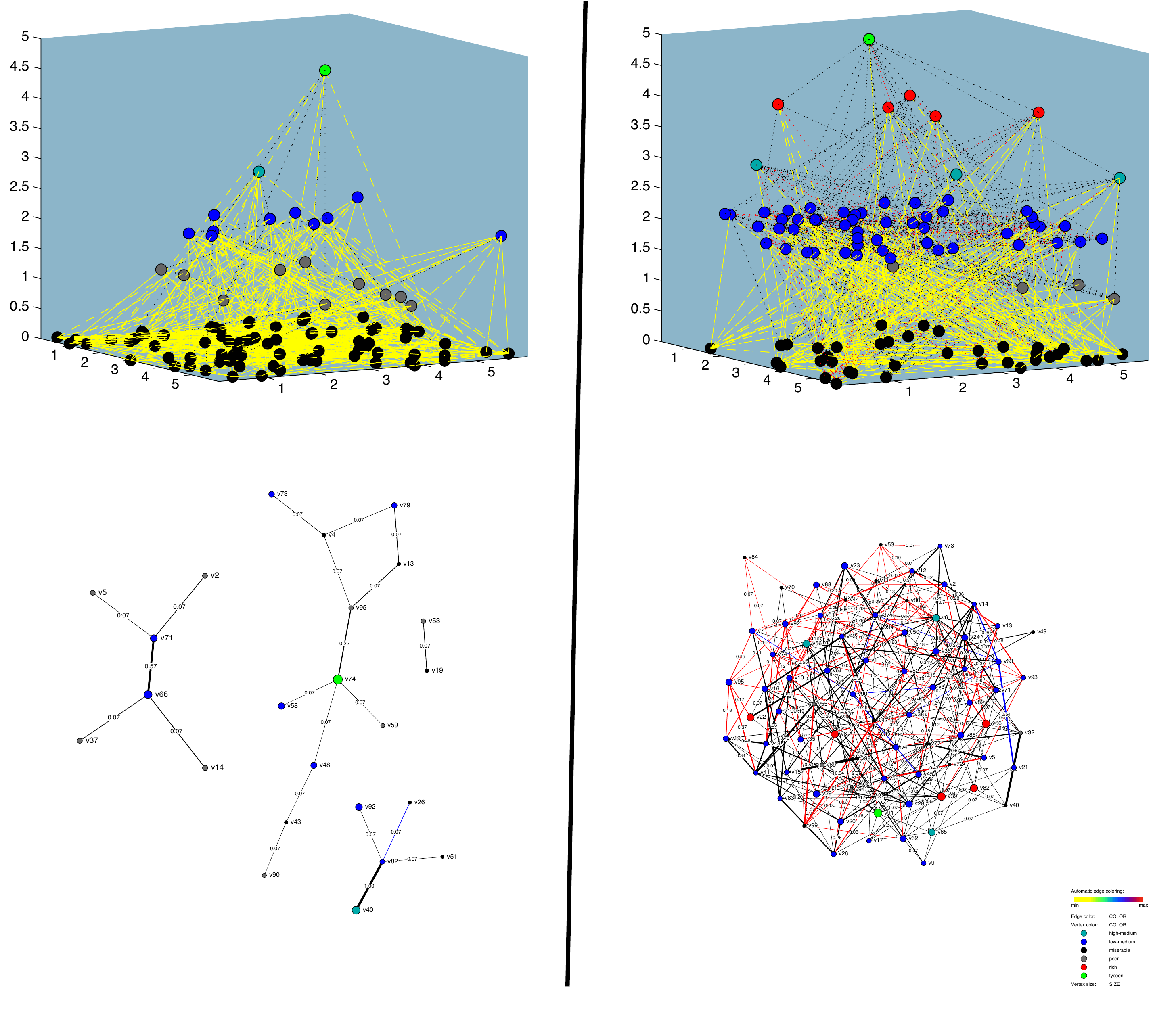}
\caption{Results of a typical network in which no trust variables are included (left hand side of the vertical black line), and including trust dynamics (right hand side). The yellow lines represent inactive links. }
 \label{fig4}
\end{center}
\end{figure}

In order to assess the effect of including trust in the transactions, we made a calculation in which no trust variables were included, that is $F_{ij}(t)=1$ and $R_i=1$, $\forall (i,t)$, and another calculation including the dynamics of the trust variables. In Fig.~\ref{fig4} we show a  representation of the network as in Fig.~\ref{fig1} and a 2D graph of the two calculations.

Here we observe that trust has two main effects: 1) it reinforces the trade network, the number of active links is increased noticeably, while in the calculation without trust (indiscriminate trading) the network is dismembered in isolated subgraphs with very few active links. 2) The wealth distribution between agents turns out to be more fair, in the sense that there are less poor people and quite a large proportion in the middle class. The backbone network, in which active trading took place, is noticeably larger and with more "black" or active interactions ($\alpha$ and $\beta \ne0$), the dramatic effect of trust is: A disconnected trading network without trust becomes a robust and connected network when trust is affecting agents in deciding transactions.

We have also observed that in the calculations with trust the state variables remain positive or zero, while in the calculations without trust a small number of agents have negative values. This means that these agents are not only poor but in debt but the number of such agents is very small, such that on average there are 160 agents with negative values in a population of 5000 agents after 200000 time steps.

\begin{figure}[ht!]
\begin{center}
\includegraphics[width=\columnwidth]{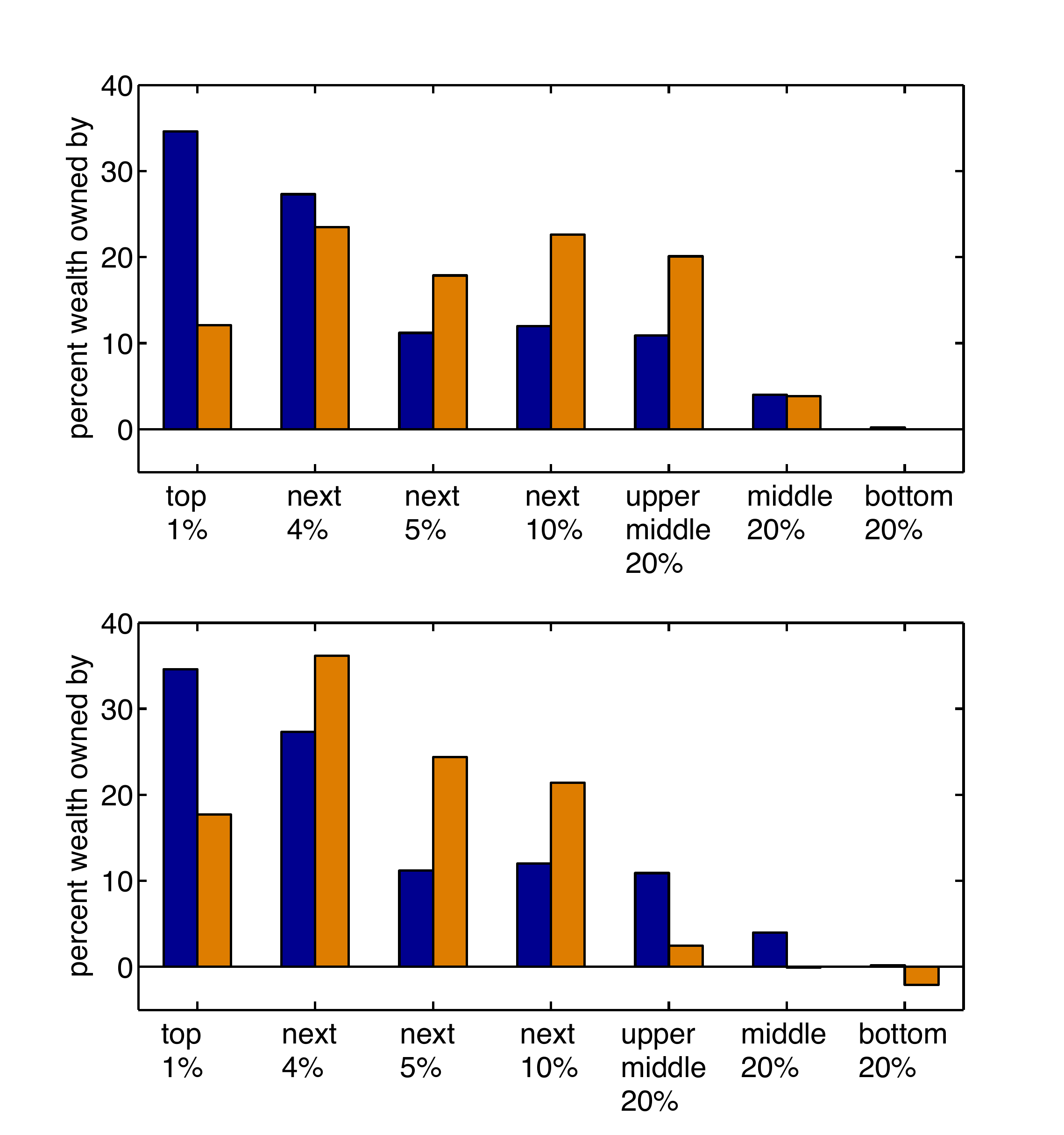}
\caption{Wealth distribution in the USA (in blue, taken from~\cite{dom}) and numerically calculated wealth distribution from our model (in yellow) including trust (top) and without including it (bottom). In both calculations the average price was $\bar{s} = 1.2$ and the spread $\mathrm{d}s = 0.4$.}
 \label{fig5a}
\end{center}
\end{figure}

As one of our main research foci we compare the distribution of wealth calculated from our model with that of the actual statistical data. For comparison we depict in Fig.~\ref{fig5a} the histograms of the actual wealth distribution in the USA (in blue), together with our results from 10 realisations of the network with 500 agents, without and with trust included.  It should be noted that our model predicts a concentration of wealth in the hands of few agents, regardless of the role of trust, although the concentration of such agents is much less pronounced than what we see in reality. In the calculation including trust the distribution amongst middle and bottom percentages agree quantitatively with the data, but we are not able to find a good match with data for the upper 40$\%$ of the wealth.  This could be due to the small size and randomness of our model network, nevertheless the trend is already noticeable.

\begin{figure}[h!]
\begin{center}
\includegraphics[width=6.5cm]{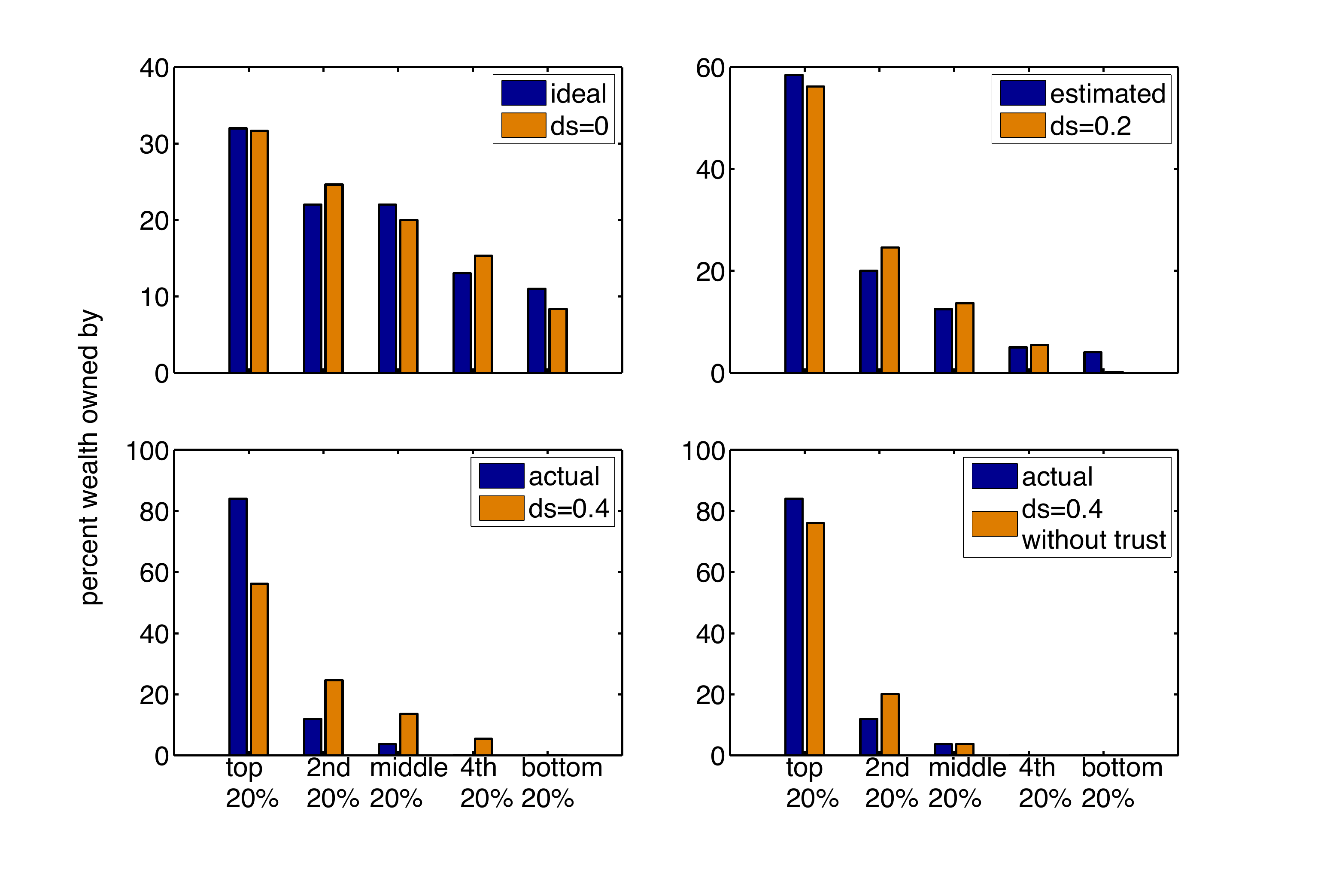}
\caption{Ideal, estimated, and actual wealth distribution in the USA (in blue, taken from~\cite{usa}) compared with numerically calculated wealth distribution from our model (in yellow) using the parameters indicated in each histogram.}
 \label{fig5b}
\end{center}
\end{figure}

In Ref.~\cite{usa} one finds an interesting exercise in which people are asked to construct wealth distribution that they consider ideal, and also an estimated distribution, based on their information, and these data are compared with actual data for wealth distribution in the USA. In Fig.~\ref{fig5b} we compare the published results with our calculations.

Here we find that our calculation results seem to compare very well with the ideal distribution when trust is included and no spread of prices is allowed. This situation probably represents a country that has strict control of unique prices for goods, set by the government or some monopoly. The estimated distribution is well reproduced with a rather small spread of prices and having trust included. However, the actual situation is somewhat disappointing, since the best fit is with a large spread of prices. In the bottom right panel we show the situation in which the network agents lack trust in their transactions thus rendering the outcome unrealistic.

\begin{figure}[ht!]
\begin{center}
\includegraphics[width=\columnwidth]{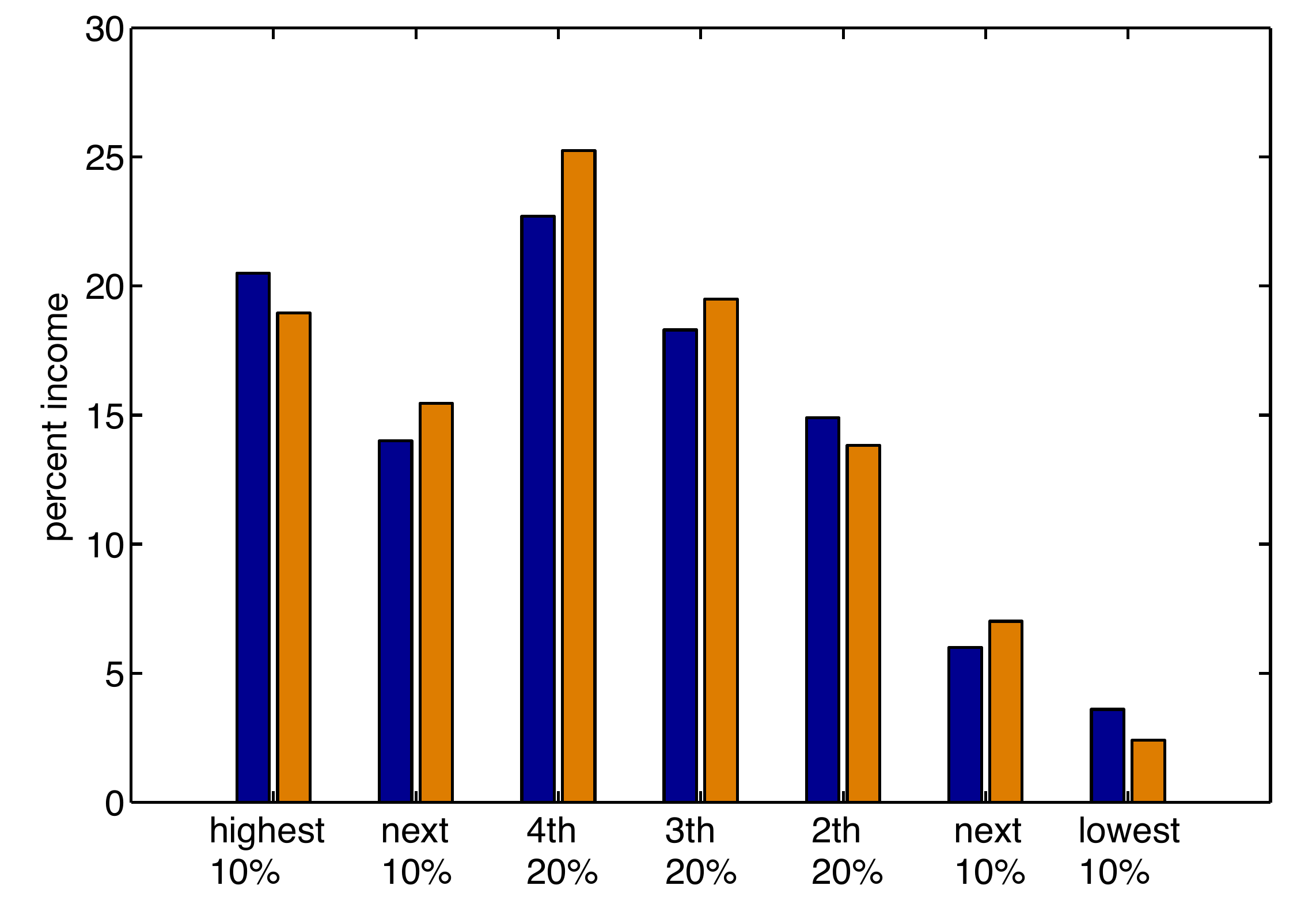}
\caption{ Actual income distribution in Denmark (1992) (in blue, taken from~\cite{din}) compared with numerically calculated wealth distribution from our model (in yellow) with trust included and having a price spread of $\mathrm{d}s = 0.02$.}
 \label{fig6}
\end{center}
\end{figure}

The situation is quite different for such egalitarian societies as Denmark, in which case we have found that for very small spread of prices and including trust our model agrees extremely well with the actual data (taken from ~\cite{din}) of the income distribution for the 1992 statistics. Using data from the money variable only, we show the comparison in Fig.~\ref{fig6}, which turns out to be very favourable.

\begin{figure}[ht!]
\begin{center}
\includegraphics[width=\columnwidth]{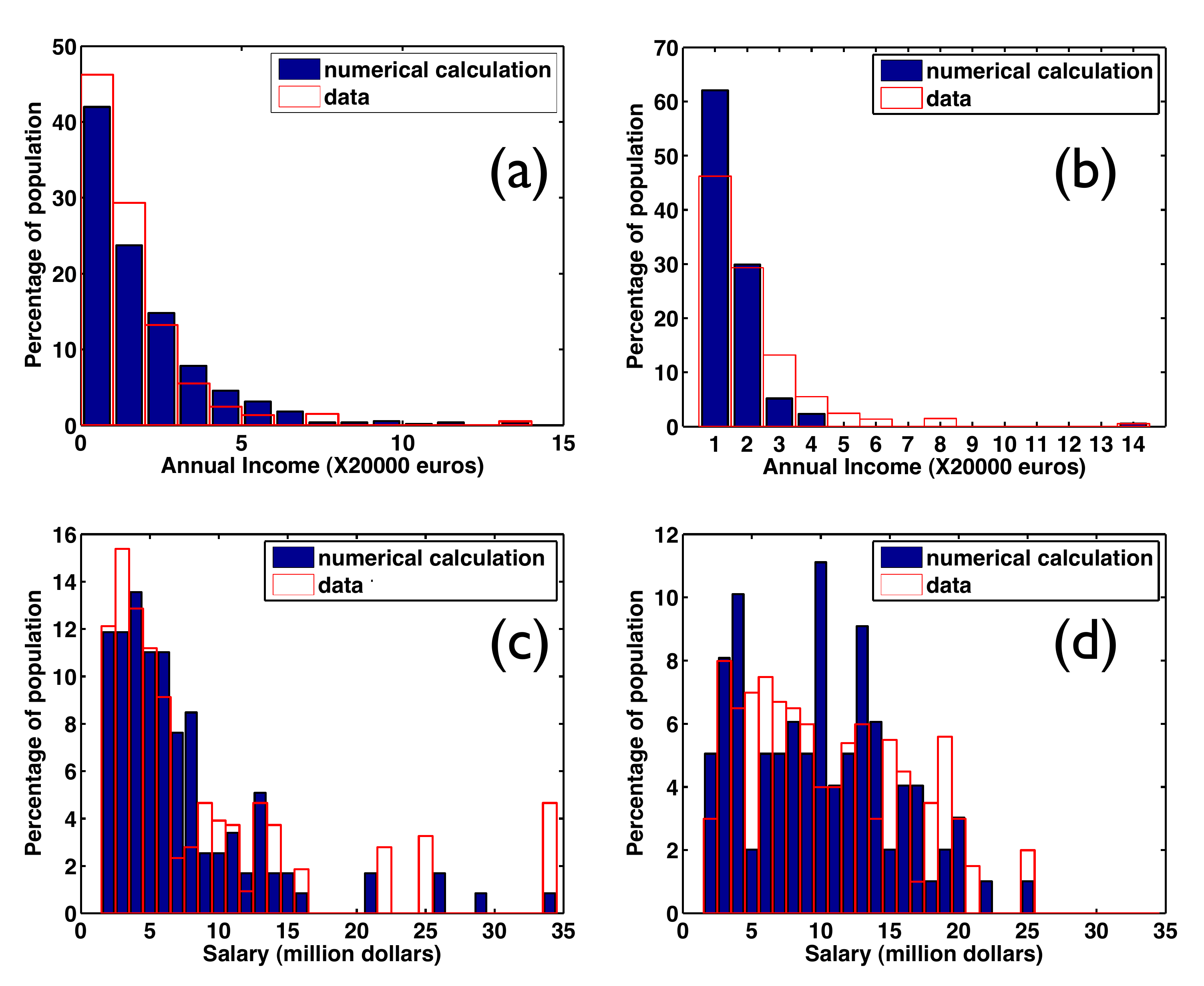}
\caption{(a) Histogram of the annual salary of european people (in red) compared with a numerical calculation without including trust in a system with 1000 agents (blue). (b) Same as (a), but the calculation was performed in a system with 250 agents. (c) Histogram of the salary of players in the NFL in 1998 (in red), compared with results from the model without including trust effects. (d) Histogram of the salary of players in the NFL in 2011 (in red), compared with results from the model including trust effects. }
 \label{fig7}
\end{center}
\end{figure}

Another maybe better way to compare our results with real statistical data is to investigate the distribution of money in different systems, since it depends on the actual mechanisms of acquiring money. For instance, there are data on the annual income of people in Europe, which ranges from 0 to millions of euros. In Fig.~\ref{fig7}(a) we compare the actual data (in red) with a histogram from our numerical calculation with 1000 agents but without including trust. In Fig.~\ref{fig7}(b) we compare the same data with the results of a calculation with 250 agents and a dispersion of prices twice as large as in (a) but once again without including trust. Both these calculations do not seem to fit with the real data, neither do averages over many realisations. \\

However, a rather different situation is encountered in a closed system as we have found out when we investigated data from the annual salaries that all the players in the NFL earned in two different years. In Fig.~\ref{fig7}(c) we show in red the distribution of salaries in 1998 and compare it with a numerical result without including trust. In Fig.~\ref{fig7}(d) we show the numerical results for the same system including trust and compare them with the NLF data from 2011. Here we can observe that both distributions fit the data fairly well, which allows us to think that trust does not play much of a role in the mechanisms of deciding salaries in a system like NFL, in which a few of the star players began to receive exaggeratedly good salaries.

\begin{figure}[ht!]
\begin{center}
\includegraphics[width=\columnwidth]{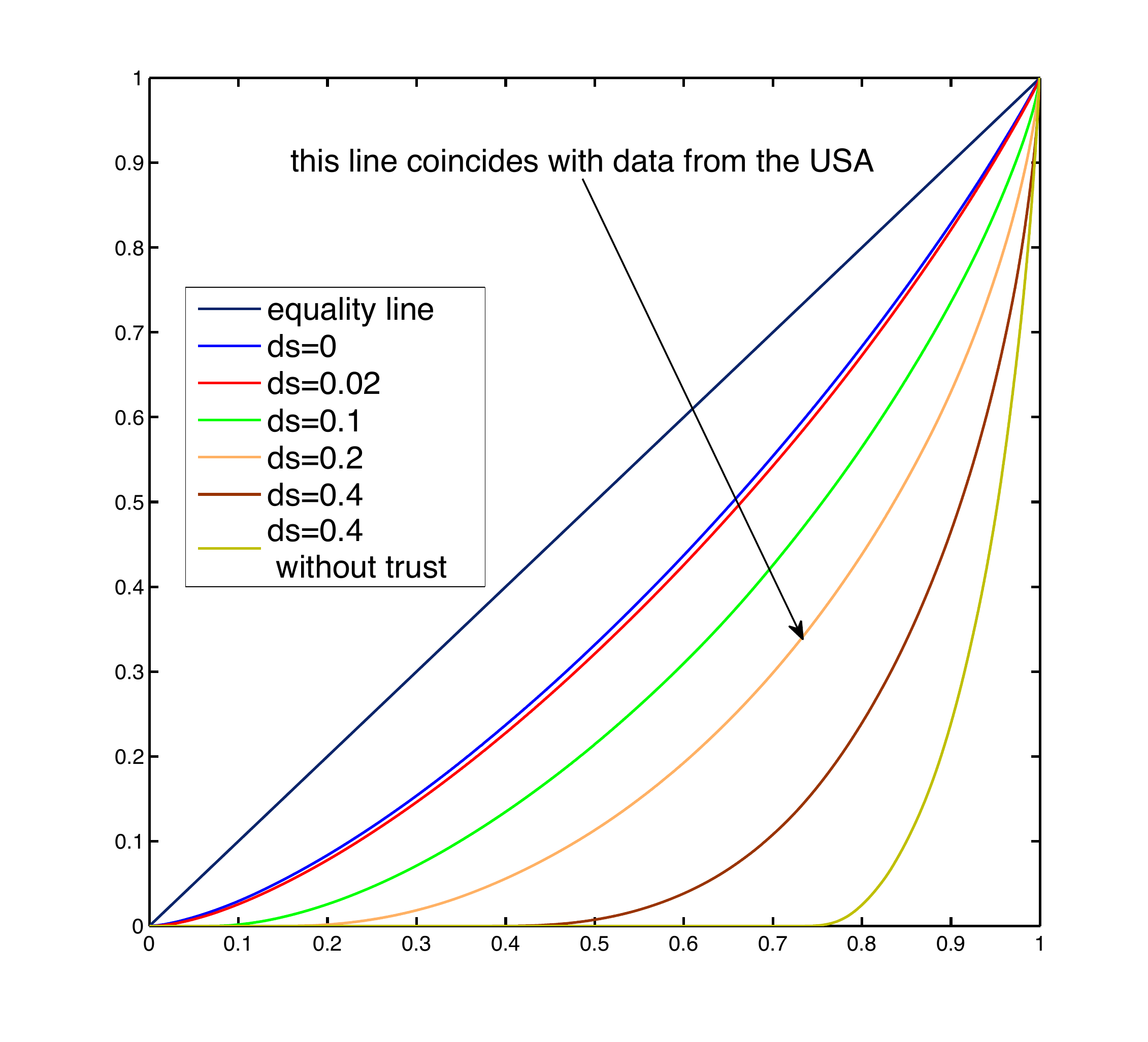}
\caption{Lorentz curves from our calculations for several values of the spread of prices. The straight black line corresponds to perfect equality. The curve indicated by the arrow matches with real data. This corroborates our claim that trading without trust makes wealth distribution skewed or unequal.}
 \label{fig8}
\end{center}
\end{figure}

In order to investigate such a situation in more detail we have adopted the approach presented by Jun-ichi Inoue et al. ~\cite{bikas}, in which they define indices to measure social inequality in various fields, including income and trading. In Fig.~\ref{fig8} we present the inequality for our model calculations. We see that these Lorentz inequality curves vary quite sensitively with the spread of prices of the goods. We also see that the real situation presents itself as rather unequal such that for the same spread of prices, the lack of trust generates more inequity. For the curve tagged with an arrow we found $g=0.5611$ and $k=0.701$, which is very similar to the indexes calculated for USA: $g=0.54-0.6$ and $k=0.69-0.71$~\cite{bikas}. We have also detected that the results do not vary much when the size of the network is increased to 1000.

\subsection{Dynamical behaviour}
\begin{figure}[ht!]
\begin{center}
\includegraphics[width=\columnwidth]{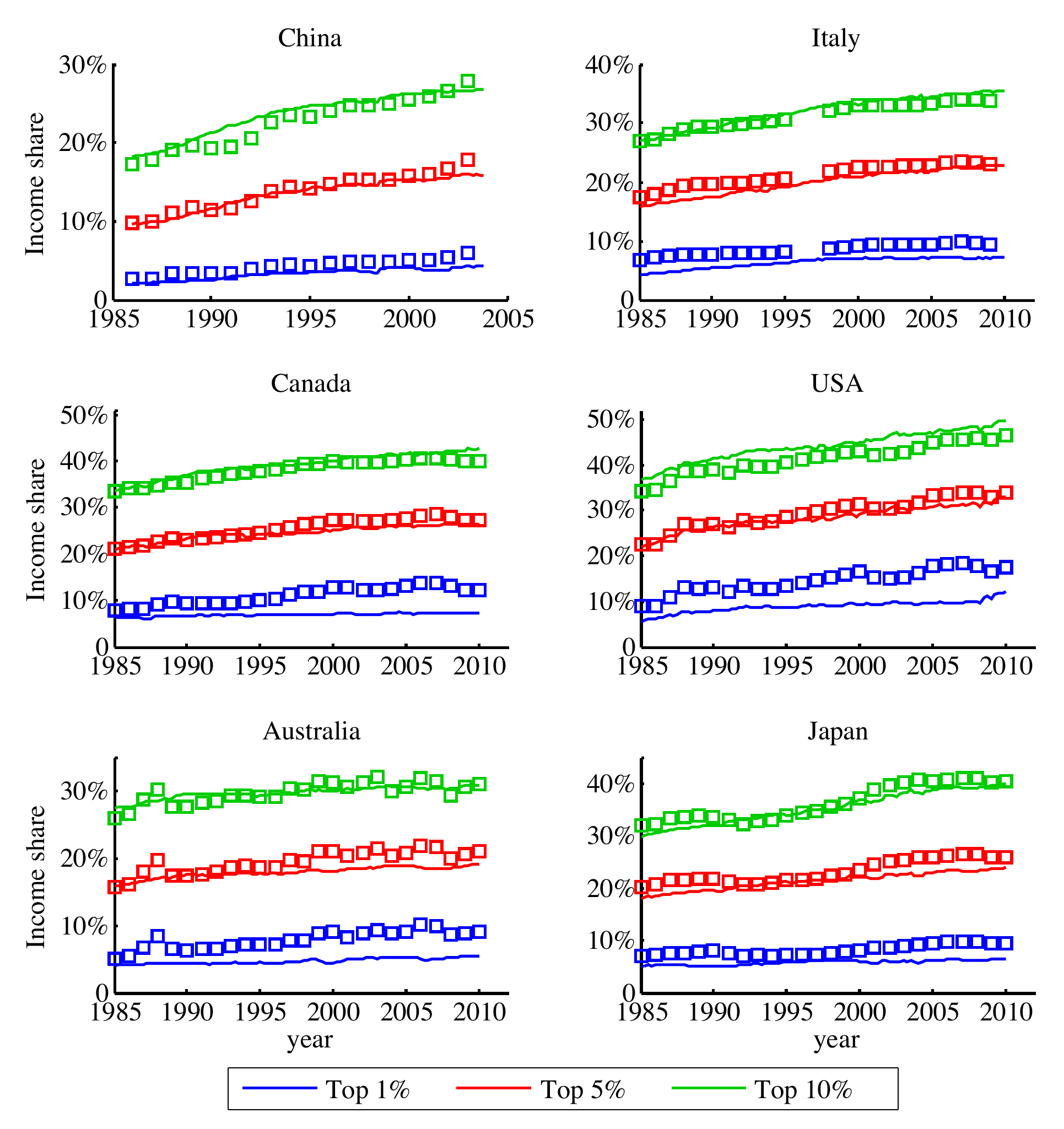}
\caption{Comparison between numerical results and actual data of the top 10 \% income share with data taken from ref.\cite{WTDB} for six selected countries.}
 \label{fig9}
\end{center}
\end{figure}

As for the dynamical behaviour of trading there are data available for the wealth share of various countries~\cite{WTDB}. In Fig.~\ref{fig9} we show the top 1, 5, and 10\% countries' income share, and compare it with the numerical calculations for a network size of 500 agents. All the calculations were set to run up to 200000 iterations, of which only half of them were selected for the comparisons. The parameters of the model $\mathrm{d}s$ and $t_0$ were varied to find the best fit.

For all the cases a value of $t_0 = 400$ in the model calculation is found to fit well with the data, except in the case of Italy, where we chose $t_0 = 600$ for the best fit.  It is interesting to notice that the case of China is the only one that fits better with the 100000 initial iterations of the calculation, and the dispersion of prices is smaller ($\mathrm{d}s = 0.15$). All the other countries are best fitted once the variables attain a final distribution, which occurs  during the last 100000 iterations. This could reflect the fact that China is a newly emergent economic power and that their rules of trading are tighter. For Japan and Australia the dispersion of prices is larger than for China ($\mathrm{d}s = 0.2$). This could be interpreted such that these countries have more free trading rules, in which case one could vary prices more widely without loosing competitiveness.

For developed countries with long history in economic traditions, $\mathrm{d}s = 0.4$ is quite large, probably reflecting the influence of many strategies to allow prices to vary without the loss of competitiveness. For instance the dispersion of gas prices nationally in the USA ranges from 1.31 to 2.37 dollars per gallon, and the prices are unevenly distributed geographically ~\cite{gas}.  Observe that USA and Canada are practicable indistinguishable, which is to be expected as they are similar and tightly linked. The case of Italy, the only European country selected, is interesting, since it is the only one in which a good fit is obtained by increasing the time scale for price changes. this seems to suggest that in Italy and similar European countries prices tend to change more slowly than in the very dynamic American economies.

Interestingly, the fit for the 1\% top population is not as good as the others, meaning that our model predicts less extremely rich people than in reality there is. The extreme social and economic inequality of the present world is most likely due to the fact that the economic alliances are not random (as in our model) and impose some trading preferences, other than the ones considered in our model. Also, our approximation of a conserved system
result in a constraint on the amount of goods or money an individual could gather, and this constraint is not present in the actual economic picture, in which money could be printed and goods could be produced and destroyed. In the real world countries cannot be considered as closed systems, although the global economy could be considered as such.

\section{Conclusions}
\label{conclusion}

In conclusion our agent-based trading model gives rise to results that overall seem to compare very favourably with the findings from the real data, even though the model takes into account only a subset of the known factors affecting trading. One striking result is the effect of trust in trading relations. First of all, it was found that trust reinforces trading transactions in such a way that the network with active links remains fully connected when trust is included, and becomes a set of disconnected graphs if it is not. Secondly, trust helps to make society more even and it is seen that the distribution of wealth is fairer when trust plays a role in trading. If trust is not included then the society seems to have a number of poor people in debt, unlike in a society with trust.

One important conclusion of this work is that even in the simplified case of having a conserved system, agreement with real data on the distribution of wealth is not only possible but also quite good. This could be interpreted indicating that including the production and deterioration of goods and money, which is essential to the idea of creation of wealth~\cite{galle}, does not seem to be a fundamental property of the economy in general. Furthermore, as far as the distribution of wealth is concerned, the fundamental issue seems to be the spread of prices, rather than the production of wealth. It also turns out that our model predicts inequality in a closed economy without production from simple rules of buying and selling between agents, which illustrates the fact that inequality can arise naturally in a rudimentary economy.

It should be noted that contrary to the general trend in classical economic models to consider various representative classes of agents (consumers or producers)~\cite{st} we have here considered a single class of heterogeneous trading agents. The behaviour emerging from such economy is an aggregate of individual decisions. Hence it seems that an unbalanced economy emerges from particular decisions of individuals~\cite{art}.

It is also important to recognise that the structure of real trading networks is more similar to a scale free networks than to random networks, which stays fixed during the  dynamic trading process. In order to test the effect of topologically changing network structure we have introduced a rewiring scheme in which the agents whose connections are not working are deleted from the network, and new agents are added following a scale free method. As a result we have found that the final network structure evolved after a long run of the dynamics resembling roughly a scale free network. It is found that the structure of the final network still depends very much on the trust variable $R$, namely if $R=1$ for all agents, then agents that become "hubs" at a certain time are very likely to be deleted, and eventually the network is fragmented into small isolated trading groups. On the other hand, if one includes $R$ in the dynamics, the hubs that are formed remain trading and the network, although changing, remains cohesive. The network that one obtains after many rewiring processes has always all the links working perfectly (black lines). However, the wealth distribution in the network is very similar to the ones reported here without rewiring, and thus we decided to leave the detailed study of the rewiring problem for the future, since the above-described rewiring scheme breaks the conservation of wealth, which is an essential assumption in our current model. Our justification to do so is that in the calculations shown here the time span for transactions is not long enough to rewire the network. \\

As the main message of this work we would like to suggest that in trade the prices seem to be the main cause of impoverishment. Wide spread of prices tends to augment the differences between poor and rich people, and the results seem much more sensitive to a change in the allowed spread of prices than to the average price. Also we conclude that trust seems to be important in regulating trading, since it is a way in which agents decide to trade preferentially amongst the agents they are linked with. In indiscriminate transactions without including trust, the network turns out to be disrupted, while with trust few links are always reinforced. In this way trust could be considered to favour the appearance of monopolies, since the agents that have a good history seem better off regardless of the high prices and quality of goods, and are thus in position of engulfing small traders.

As a final remark we could conclude that our simple model conserving the amount of goods and money is able to reproduce some salient features found in trading networks, with the additional advantage that we could fairly easily add new features to the model and analyse them in depth.

\section*{Acknowledgments}
RB, ERG, and TG would like to acknowledge financial support from Conacyt (Mexico) through project 179616 and KK from Academy of Finland through project 276439. We are grateful to Prof. Larissa Adler, whose original ideas were the primary source of inspiration for this work. We are also grateful to  Prof. Juan M. Hern\'andez for careful reading of the manuscript and valuable comments and criticisms.



\begin{thebibliography}{10}

\bibitem{gamb} Gambetta D (ed.), {\it Trust: Making and Breaking Cooperative Relations}. Oxford University Press (1988).

\bibitem{dunbar} Dunbar R, {\it How Many Friends Does One Person Need? Dunbar's Number and Other Evolutionary Quirks}, Faber and Faber (2010) ISBN: 9780571253425.

\bibitem{sinha} Sinha S, {\it Stochastic Maps, Wealth Distribution in Random Asset Exchange Models and the Marginal Utility of Relative Wealth} Phys. Scripta T{\bf 106}, 59 (2003).

\bibitem{wd} Berman Y, Shapira Y, Ben-Jacob E {\it Modeling the Origin and Possible Control of the Wealth Inequality Surge}. PLoS ONE 10(6): e0130181. (2015) doi:10.1371/journal.pone.0130181
  

\bibitem{bikas2007} Chatterjee A, Chakrabarti BK, {\it Kinetic exchange models for income and wealth distributions},
Eur. Phys. J. B {\bf 60}, 135-149 (2007) doi:10.1140/epjb/e2007-00343-8

\bibitem{wang} Wang Y, Qiu H, {\it The velocity of money in a life-cycle model}, Physica A {\bf 353}, 493-500 (2005), ISSN 0378-4371, doi:10.1016/j.physa.2005.01.053
  
\bibitem{bikas2015} Chakraborti A, Challet D, Chatterjee A, Marsili M, Zhang Y, Chakrabarti BK, {\it Statistical mechanics of competitive resource allocation using agent-based models}, Physics Reports, {\bf 552}, 1-25 (2015), ISSN 0370-1573, doi:10.1016/j.physrep.2014.09.006
  
\bibitem{sornette} Sornette D, {\it Physics and financial economics (1776-2014): puzzles, Ising and agent-based models} Rep. Prog. Phys. {\bf 77(6)}, 062001 (2014)

\bibitem{yuqing2007} Yuqing H, {\it Income distribution: Boltzmann analysis and its extension}. Physica A {\bf 377}, 230 (2007) ISSN 0378-4371, doi:10.1016/j.physa.2006.11.009

\bibitem{sor} Sornette D, {\it Critical market crashes}, Physics Reports, 1 {\bf 378} 1-98 (2003), ISSN: 0370-1573, doi:10.1016/S0370-1573(02)00634-8.
  
\bibitem{usa} Norton MI, Ariely D, {\it Building a Better America--One Wealth Quintile at a Time}, Perspectives on Psychological Science, 6 {\bf 1} (2011), doi:10.1177/1745691610393524

\bibitem{din} Nations Encyclopedia, {\it Denmark - Poverty and wealth}. \\ \url{http://www.nationsencyclopedia.com/economies/Europe/Denmark-POVERTY-AND-WEALTH.html}

\bibitem{dom} Domhoff GW, {\it Wealth, Income, and Power} UC-Santa Barbara Sociology Department. \url{http://www2.ucsc.edu/whorulesamerica/power/wealth.html}

\bibitem{bikas} Inoue J, Ghosh A, Chatterjee A, Chakrabarti BK, {\it Measuring social inequality with quantitative methodology: Analytical estimates and empirical data analysis by Gini and indices}, Physica A {\bf 429(1)} 184-204 (2015) ISSN 0378-4371, doi:10.1016/j.physa.2015.01.082

\bibitem{WTDB} The world top income database. \url{http://topincomes.parisschoolofeconomics.eu/}

\bibitem{gas} GasBuddy.com  \url{http://www.gasbuddy.com}

\bibitem{galle} Gallegati M, Keen S, Lux T, Ormerod P, {\it Worrying trends in econophysics}, Physica A {\bf 370(1)} 1-6 (2006) ISSN 0378-4371, doi:10.1016/j.physa.2006.04.029
  
\bibitem{art} Arthur WB, {\it Out-of-Equilibrium Economics and Agent-Based Modeling}, Handbook of Computational Economics, Elsevier {\bf 2} 1551-1564 (2006) ISBN 9780444512536, doi:10.1016/S1574-0021(05)02032-0
  
\bibitem{st} Stiglitz JE, {\it Distribution of income and wealth Among individuals}, Econometrica 37, 382-397 (1969) doi:10.2307/1912788 

\end{thebibliography}
\end{document}